\documentclass{PoS}
\usepackage[center]{subfigure}
\title{Neutrino Interactions\\ and Long-Baseline Experiments}

\ShortTitle{Neutrino Interactions}

\author{Ulrich Mosel\\
        Institut fuer Theoretische Physik, Universitaet Giessen, 35392 Giessen, Germany\\
        E-mail: \email{mosel@physik.uni-giessen.de}}

\abstract{The extraction of neutrino mixing parameters and the CP-violating phase requires knowledge of the neutrino energy. This energy must be reconstructed from the final state of a neutrino-nucleus reaction since all long-baseline experiments use nuclear targets. This reconstruction requires detailed knowledge of the neutrino reactions with bound nucleons and of the final state interactions of hadrons with the nuclear environment. Quantum-kinetic transport theory can be used to build an event generator for this reconstruction that takes basic nuclear properties, such as binding, into account. Some examples are discussed that show the effects of nuclear interactions on observables in long-baseline experiments.}

\FullConference{38th International Conference on High Energy Physics\\
		3-10 August 2016\\
		Chicago, USA}

\begin{document}

\section{Introduction}
Long-baseline neutrino experiments search for oscillation patterns by comparing neutrino-target interaction event rates at a near and a far detector. Any differences between these event rates depend on the neutrino mixing parameters and phases. In addition, they also depend on properties of the near and far detectors and on the energy of the incoming neutrino. Unlike in any other experiment in nuclear or particle physics the latter is not known, but must be reconstructed from the observed parts of the final state phase space of a neutrino-nucleus reaction. This requires a reliable description both of the initial interaction and the further final-state development.

\section{Model}
Neutrino event generators are needed to follow the nuclear reaction from its initial interaction all the way to the final state. The most widely used generators describe the nuclear ground state as an ensemble of unbound nucleons, with a momentum distribution of the relativistic Fermi gas. Final state interactions are described by a Monte Carlo sampling of nuclear collisions and free motion in between collisions. Standard neutrino generators thus neglect nuclear binding, and the corresponding nucleon potentials, both essential properties of nuclei, from the outset.

Nuclear transport theory, on the contrary, is able to provide a description of neutrino-nucleus interactions while taking the binding of nuclei and the in-medium transport of hadrons into account. Transport theory is widely being applied to other nuclear physics reactions, and to neutrino transport. It can be strictly derived from quantum-kinetic theory and can take care of off-shell processes and transport between collisions \cite{Buss:2011mx}. Furthermore, from the outset it is relativistically covariant. The results shown in this short paper were all obtained with the GiBUU implementation of transport theory,  which has been widely tested on a broad class of nuclear reactions, such as $A+A$, $p + A$, $\pi + A$ and $\gamma + A$. It gives not just inclusive cross sections, but also the full final state, i.e.\ the four-vectors of all final-state particles. Details on the derivation of the transport equations and on its special GiBUU implementation can be found in Ref.\ \cite{Buss:2011mx} while details on recent improvements for the description of the ground state and of 2p2h interactions are contained in Ref.\ \cite{Gallmeister:2016dnq}.

\section{Results}

\subsection{Electron Scattering}
\begin{figure}
\centering
\includegraphics[width=.45\textwidth,angle=-90]{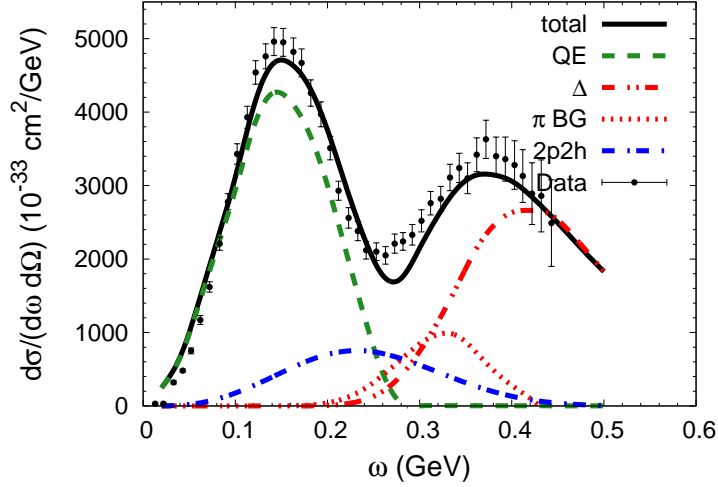}
\caption{Inclusive cross section for scattering of electrons on carbon at 560 MeV and 60 Deg ($Q^2 = 0.24$ GeV$^2$ at the QE peak), obtained with a free $\Delta$ spectral function. The leftmost dashed curve gives the contribution from true QE scattering, the dash-dotted curve that from 2p-2h processes, the dashed-dotted-dotted curve that from $\Delta$ excitation and the dotted curve that from pion background terms. From \cite{Gallmeister:2016dnq}.}
\label{fig:eC}
\end{figure}

Inclusive cross sections for $(e,A)$ reactions, which differ from neutrino-induced ones only by the presence of an axial coupling, provide a necessary test for the quality of the theoretical description of a ($\nu,A$) reaction. They are dominated by the first, initial reaction of the incoming lepton with the nucleus. Final-state interactions enter only in so far as they influence the phase-space of the outgoing hadrons in the first interaction and thus the interaction rate.  An example for the quality that can be achieved with GiBUU in the description of inclusive electron scattering cross sections is shown in Fig.\ \ref{fig:eC}. The figure not only shows the total cross section, but also the different reaction channels contributing to it. In the kinematical regime of this figure quasielastic scattering (QE), 2p2h interactions and pion production, both through the $\Delta$ resonance and through non-resonant interactions, contribute and partly overlap. This illustrates the need to describe various different reaction mechanisms.

\subsection{Neutrino Scattering}
When neutrinos interact with nuclei axial couplings and form factors come into play. While the QE scattering process is determined by only one such form factor, pion production as the first inelastic process is governed by three axial form factors. All of these form factors carry much larger uncertainties than the corresponding vector form factors that can be determined quite accurately in electron-scattering experiments. 

\subsubsection{Inclusive Cross Sections}
\begin{figure}[ht]
\begin{center}
\subfigure{
\includegraphics[width=0.6\linewidth,angle=0]{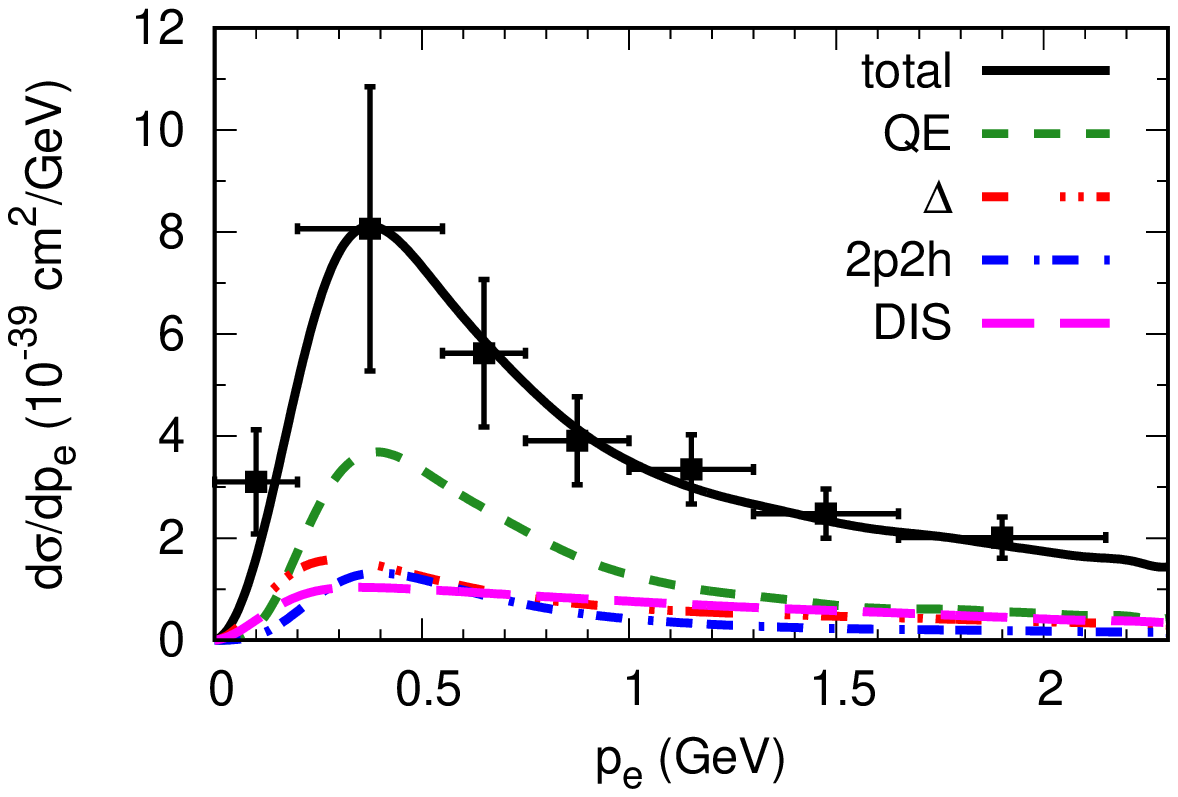}
}
\subfigure{
\includegraphics[width=0.6\linewidth,angle=0]{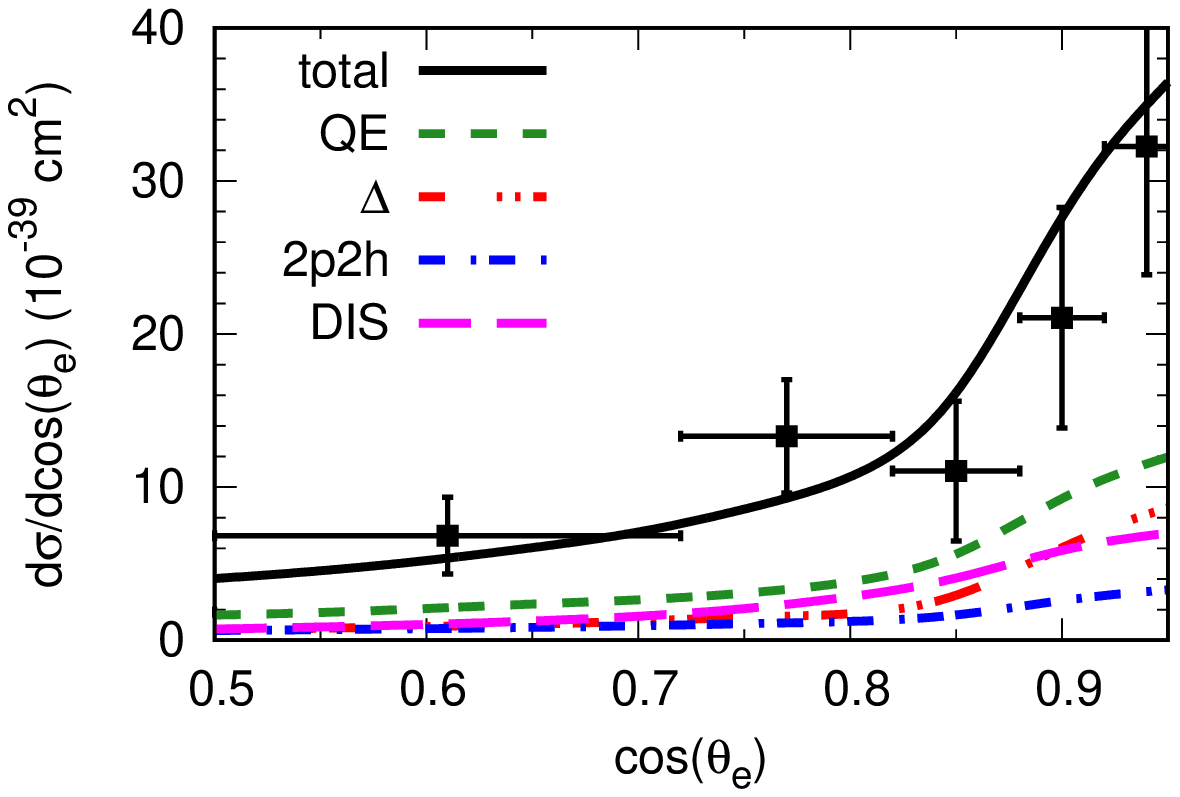}
}
\end{center}
\caption{Momentum (top panel) and angular distribution (bottom panel) of fully inclusive events for an electron-neutrino beam on a C target in the T2K near detector. The solid curve gives the sum of all contributions; the contributions of some dominant reaction channels are explicitly indicated in the figure: QE (green, dashed), $\Delta$ excitation (red, dash-dot-dotted), 2p2h (blue, dash-dotted) and DIS (magenta, long-dashed). From \cite{Gallmeister:2016dnq}.}
\label{fig:T2K_ND_p-ecost}
\end{figure}

The near detector of the T2K experiment has given interesting information on the neutrino-nucleus reactions. Fig.\ \ref{fig:T2K_ND_p-ecost} shows cross sections measured with this detector for the interactions of an electron-neutrino beam. It is interesting to see that even at the relatively low average energy (approximately 0.7 GeV) of the T2K neutrino beam many different processes contribute. While QE scattering is dominant, it accounts only for about one half of the total inclusive cross section at the maximum of the momentum distribution (upper figure in Fig.\ \ref{fig:T2K_ND_p-ecost}). The remainder is made up by $\Delta$ resonance  excitation, 2p2h excitations and even deep inelastic scattering (DIS). The last process contributes because the distribution of incoming neutrino energies has a high-energy tail and the cross section for DIS is proportional to the neutrino energy.

\begin{figure} 
	\includegraphics[width=.8\textwidth]{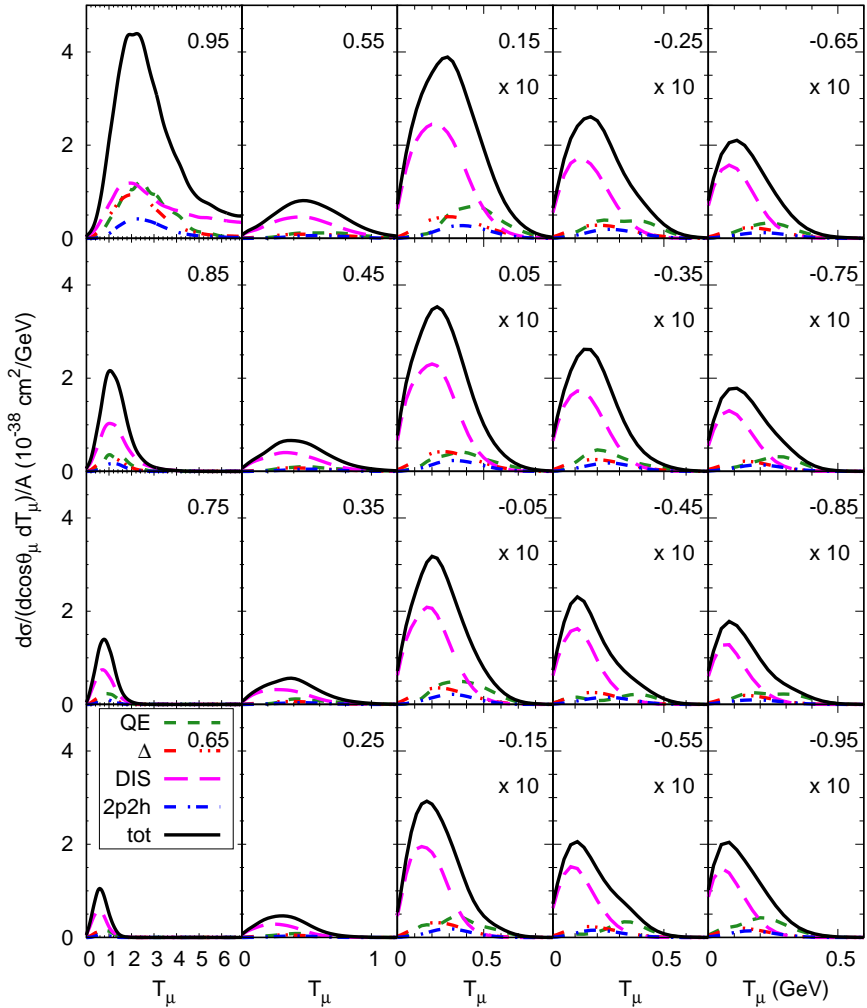}
	\caption{Fully inclusive double differential cross section per nucleon for the muon neutrino beam from the LBNF. The numbers in the upper right corner give the cosine of the muon scattering angle while $T_\mu$ is the muon's kinetic energy in GeV. For angles with $\cos \theta \le 0.15$ the cross section scale has been changed by a factor of 10 to make the composition at backward angles more visible. The dominant reaction contributions are denoted as given in the figure. Taken from \cite{Gallmeister:2016dnq}}
	\label{fig:DUNE}
\end{figure}

At the higher energies of the planned DUNE experiment DIS is expected to become more dominant. This is indeed seen in Fig.\ \ref{fig:DUNE} that shows the double-differential cross section expected at the LBNF near detector. Here DIS is the strongest subprocess. Only at very forward angles (top left subfigure)
QE and  $\Delta$ excitation are of comparable strength; 2p2h is relatively small at all angles. 

\subsubsection{Exclusive Cross Sections and Energy Reconstruction}
For any extraction of the neutrino mixing parameters and phases the incoming-neutrino's energy has to be known. Since only energy distributions can reasonably well be assigned an event-by-event energy reconstruction from final-state properties is necessary. 

Most high-energy long-baseline experiments rely for this reconstruction on a calorimetric method in which the energies of the outgoing particles are measured. In an ideal detector their total energy equals that of the incoming neutrino. In an unideal world, however, detectors have acceptances and thresholds so that a generator has to be used to compute backwards from the observed parts of the final state phase-space to the initial state of a nucleus in its groundstate and an incoming neutrino.

Another method, mostly used for lower-energy experiments, is a QE-based method. If an elementary process could be uniquely identified as being QE-scattering, then the energy and angle of the outgoing lepton determine the incoming neutrino energy. Since in a nuclear target the nucleons are Fermi-moving there will necessarily be a smearing over these energies. In addition, there is the problem to identify single particle QE scattering because pion production events in which the pions get reabsorbed are experimentally indistinguishable from true QE scattering events. In Ref.\ \cite{Mosel:2013fxa} it has been shown that this crucial event identification can be greatly improved if events with zero pions and one (and only one) proton are selected; in addition there could be $X$ neutrons. The restriction to one proton events implicitly selects peripheral events in the surface of the target nucleus and this minimizes final state interactions.

\begin{figure}
\centering
\includegraphics[angle=-90,width=0.6\textwidth]{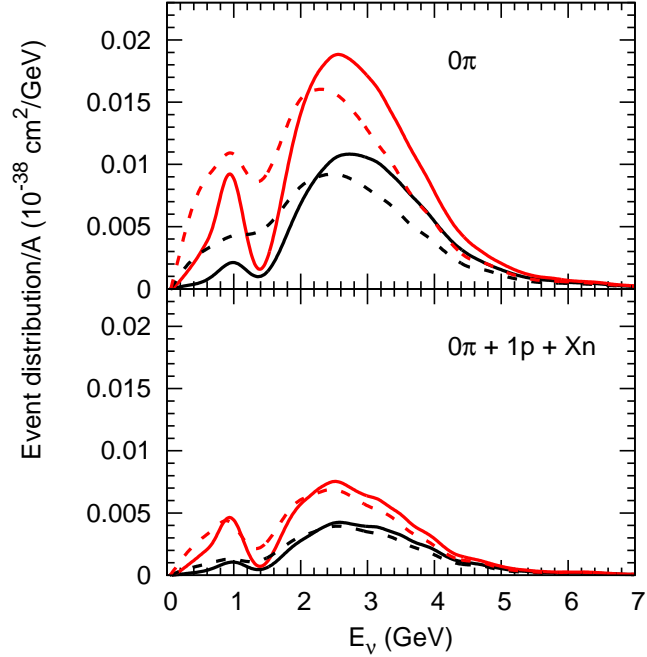}
\caption{(Color online) Event distributions per nucleon for electron appearance for $\delta_{CP}=+ \pi/2$ and $\delta_{CP}=-\pi/2$ (upper red and lower black curves, resp.), both for true (solid) and reconstructed (dashed) energies for the DUNE flux. The upper part gives the results for 0 pion events, the lower that for events with 0 pions, 1 proton and X neutrons. The reconstruction was done with the QE-based method. Taken from \cite{Mosel:2013fxa}.} \label{fig:deltaCP}
\end{figure}

Fig.\ \ref{fig:deltaCP} shows the oscillation signal in the DUNE experiment and its variance with a $CP$ violating phase. The true-energy curves have been plotted as a function of the incoming energy as it enters into GiBUU, whereas the reconstructed-energy curves have been plotted versus a reconstructed energy. If the neutrino energy is reconstructed from an event sample with the only requirement that it contains zero pions, then the oscillation signal is severely smeared and shifted towards lower energies, as shown by the dashed curves in the upper part of the figure. On the other hand, if the additional requirement of one and only one proton is used, then the oscillation signals lie much closer to each other and are within the required energy uncertainty of about 100 MeV (lower part of figure).

\end{document}